\documentstyle[11pt]{article}
\begin{document}

\begin{flushright}
{hep-ph/0005xxx}\\

\end{flushright}
\vspace{1cm}
\begin{center}
{\Large \bf  Neutrino Oscillations in Supersymmetry without
Lepton number conservation and R-parity}
\vspace{.2cm}
\end{center}
\vspace{1cm}
\begin{center}
{Tai-Fu Feng$^{1,2,3}$ \hspace{0.5cm}
Xue-Qian Li$^{1,2,3}$}\\
\vspace{.5cm}

{$^1$CCAST (World Laboratory), P.O.Box 8730,
Beijing 100080, China}\\
{$^2$Department of Physics, Nankai
University, Tianjin 300070, China}\\
{$^3$Institute of Theoretical Physics, Academia Sinica, P.O. Box
2735, Beijing 100080, China}\\

\vspace{.5cm}

\end{center}
\hspace{3in}

\begin{center}
\begin{minipage}{11cm}
{\large\bf Abstract}

{\small With the on-shell renormalization scheme,
we discuss  neutrino masses
up to one-loop approximation in the Supersymmetry without lepton
number conservation and R-parity. Ii is shown that in this
model with experimentally allowed parameters,
$\Delta m^2_{32},\;\Delta m^2_{12}$ and the mixing angles
$|\sin\theta{23}|,\;|\sin\theta_{12}|$ which
are consistent with the present observation
values can be produced.
We find that small neutrino mass ($\leq$ 1 eV)
sets a loose constraint on the R-parity violation parameters in
the soft breaking terms.}
\end{minipage}
\end{center}

\vspace{4mm}
\begin{center}
{\large{\bf PACS numbers:} 12.10.Ff, 12.60.Jv, 13.10+q.} \\
\end{center}

{\large{\bf Keywords:} Supersymmetry, R-parity, Neutrino, Oscillations.}

\baselineskip 22pt

\section{Introduction}

Nonzero neutrino mass\cite{osci} implies that new physics beyond the
Minimal Standard
Model (MSM) must exist. During the last two decades, the minimal supersymmetric
extension of the minimal standard model (MSSM)\cite{wess} has been studied
carefully. In the supersymmetric model, the renormalizable Lagrangian
can include the terms without lepton (L) number and/or baryon (B) number conservation.
Usually, we can
remove such terms by imposing the R-parity symmetry, with $R=(-1)^{3B+2L+2S}$
(S is the spin of the particle). By this definition, the R-value of the
SM particles is +1 and that of the supersymmetric partners is $-1$. Alternatively, we
can dismiss the R-parity conservation and retain the B or L violation terms
in the Lagrangian. Because the proton decay data have set a very strict
constraint on the B-violation, we presume the B-conservation  and
only keep an L-number violating interaction in the Lagrangian.
In the framework of the supersymmetric model without R-parity, some
interesting phenomenological problems such as $\mu\rightarrow
e\gamma$, neutrino mass and etc. were discussed in \cite{add}.
Recently, the supersymmetric model with L-number  violation has
been employed by many authors to explain the present atmospheric
neutrino experiments \cite{rpb,rpren}.
The attractive point of the theory is that
one generation neutrino can acquire mass through its mixing with
gauginos and higgsinos \cite{remp} and the other two-generation neutrinos
can acquire their masses through loop corrections\cite{grosshaber}.
This idea urges people
to investigate the mechanism which may induce neutrino masses
\cite{rpvneu}. If in the parameter space one imposes a relation
$\epsilon_2\cos\theta_v+\mu\sin\theta_v\approx 0$ where
$\cos\theta_v=v_d/v_L$ and $v_d,v_L$ are the vacuum expectation
values of the d-type Higgs and scalar lepton,
$\epsilon_2,\mu$ are the R-parity conserving and violating parameters
respectively, he can expect small $\nu_{\tau}$ mass due to the
generalized SUSY see-saw mechanism. Recently,
the authors of \cite{Davidson} have
computed the one-loop correction to neutrino masses in the
$\overline{MS}$-scheme, they have given a very strict constraint on the
parameter space. In this work, we carry out a complete calculation
up to one-loop order in the on-shell renormalization scheme and
investigate the constraint on the parameter space
by the neutrino oscillation data.

It is well known that renormalization is carried out to remove the
ultraviolet divergence which
appears in the loop calculations. At present, two renormalization schemes
are often adopted in the literature. The first is the minimal subtraction
(MS) and modified minimal subtraction ($\overline{MS}$) schemes, these schemes
are often applied
in the QCD-calculations. The reason is that quarks and gluons are
confined inside hadrons, no free quark or gluon can be directly
observed in experiments, thus their physical masses are not
measurable quantities.
The second is the on-mass-shell subtraction
scheme\cite{denner}, that is often used in the EW-process calculations.
The advantage of the on-shell scheme is that all parameters have
clear physical meaning and can be measured directly in
experiments. But in general, practical calculations
in this scheme are more complicated than  in the
$\overline{MS}$-scheme. In this paper, we would adopt the on-shell scheme
for its advantage. Furthermore, we would perform our calculation in
a strict form because any inappropriate approximation may impede
us to understand  why neutrino mass is so small,
unless  we have a very strong reason to take any approximation.
Our main results can be summarized as follows:
\begin{itemize}
\item Using the on-shell renormalization scheme, we find that the loop
corrections to the mass matrix elements of neutrino-neutralino are decreasing
when masses of the scalar particles turn larger.
\item The neutrino oscillation data impose a loose constraint on the parameters
of the soft breaking terms.
\end{itemize}

An extensive analysis of the neutrino oscillations in terms of the
supersymmetric model with bilinear R-parity violation is made in \cite{news}.
Although a different method is adopted in this work, our results
qualitatively coincide with those of \cite{news}.

The paper is organized as follows: in Sec.\ref{on-shell}, we
present the basic ingredients of the supersymmetric theory without
lepton number conservation and R-parity. Using the on-shell
renormalization scheme, the one-loop corrections to the neutrino
masses are also included in the section. In  Sec.\ref{model},
we first discuss the one-loop corrections to the neutrino mass
with only $\tau$-number violation. Then, we generalize the
discussion with three-generation lepton number violation. We close
this paper with conclusions and discussions in the last section. Most
of the technical details are omitted in the text and then
collected in the appendices.

\section{The Lagrangian and the on-shell renormalization  scheme\label{on-shell}}

In the supersymmetric extension of the standard model without the
lepton-number conservation, the down-type Higgs superfield and lepton superfield have
the same gauge quantum number, we can combine them into a vector
$\hat{L}_{J}$=($\hat{H}_{d}$, $\hat{L}_i$) with J=0, 1, 2, 3. Using this notation,
the superpotential can be written as
\begin{eqnarray}
&&W=\mu^{J}\hat{H}\hat{L}_J+\lambda^{JKl}\hat{L}_J\hat{L}_{K}E_l^c
+\lambda_d^{Jpq}\hat{L}_J\hat{Q}_p\hat{D}_q^c+h_{u}^{pq}\hat{H}_u
\hat{Q}_p\hat{U}_q^c,
\label{superpotential}
\end{eqnarray}
where $\lambda^{JKl}=-\lambda^{KJl}$. The parameters $\mu^0$,
$\lambda^{0kl}$ and $\lambda_d^{0pq}$ are the R-parity conserving
coupling constants. The other parameters such as $\mu^i$, $\lambda^{kml}$
and $\lambda_d^{kpq}$ represent the R-parity violation coupling
constants. Here we  use the subscripts  $i$, $j$, $k$, $l$,
$\cdots$ to represent the generation indices of the leptons and quarks.

In order to break  supersymmetry, the following soft terms
should be introduced:
\begin{eqnarray}
&&V_{soft}=\frac{m_{H_u}^2}{2}H_{u}^\dagger H_u+\frac{1}{2}L^{J\dagger}
(m_{L}^2)_{JK}L^K+B^JH_uL_J\nonumber \\
&&\hspace{1.5cm}+A_uH_uQU^c+A_d^JL_JQD^c+A_e^{JK}L_JL_KE^c+h.c.
\label{soft}
\end{eqnarray}
Note, here we do not invoke the M-SUGRA scenario and absorb the superpotential
parameters into the soft breaking parameters. In terms of $W$ and
$V_{soft}$ given in
Eq.(\ref{superpotential}) and Eq.(\ref{soft}), we can compute the loop
corrections to the neutrino masses. In this paper, we will perform
our analysis in the mass basis, which is independent of the U(4)-rotation
among the superfields $\hat{L}_J$. The relevant Feynman rules can be found in
\cite{CFeng,news} and we only cite those contents of the references which are
necessary to the calculations of this work.

Now, we compute the one-loop corrections to the neutrino masses in
the supersymmetric extension without R-parity. The general form of
the self-energy for $\kappa_i^0-\kappa_j^0$ can be written
as
\begin{eqnarray}
&&\Sigma(k)_{ij}=c_{ij}m_{j}\omega_-+d_{ij}m_{i}\omega_++e_{ij}
/\!\!\! k\omega_-+f_{ij}/\!\!\! k \omega_+.
\label{form}
\end{eqnarray}
When one external leg of the self-energy is neutrino,  $k^2\ll m_0^2$
with $m_0$ being the mass of the heaviest internal-particle and we can
write  $c_{ij}$, $d_{ij}$, $e_{ij}$ and $f_{ij}$ as
expressions of $k^2$\cite{yao}:
\begin{eqnarray}
&&c_{ij}=c_{ij}^0+k^2c_{ij}^1, \!\! \nonumber\\
&&d_{ij}=d_{ij}^0+k^2d_{ij}^1, \!\! \nonumber\\
&&e_{ij}=e_{ij}^0+k^2e_{ij}^1, \!\! \nonumber\\
&&f_{ij}=f_{ij}^0+k^2f_{ij}^1. \!\!
\label{series}
\end{eqnarray}
$\Sigma_{ij}$'s are renormalized by adding counter-terms and the renormalized
$\Sigma_{ij}^{REN}$ are written as:
\begin{equation}
\Sigma_{ij}^{REN}(k)=\Sigma_{ij}(k)+\Big(c_{ij}^*m_{j}\omega_-+d_{ij}^*m_i
\omega_++e_{ij}^*/\!\!\!k\omega_-+f_{ij}^*/\!\!\!k\omega_+\Big),
\label{counter}
\end{equation}
where the quantities with * are the counter parts.
In the on-shell renormalization scheme they are determined by the mass-shell
conditions
\begin{eqnarray}
\Sigma^{REN}_{ij}(k)u_i(k)|_{k^2=m_i^2}=0,\\\nonumber
\bar{u}_{j}(k)\Sigma^{REN}_{ij}(k)|_{k^2=m_j^2}=0,
\label{shell}
\end{eqnarray}
which yield  solutions:
\begin{eqnarray}
&&c_{ij}^*=-c_{ij}^0+m_i^2d_{ij}^1+m_i^2e_{ij}^1+m_im_jf_{ij}^1,\nonumber \\
&&d_{ij}^*=-d_{ij}^0+m_j^2c_{ij}^1+m_j^2e_{ij}^1+m_im_jf_{ij}^1,\nonumber \\
&&e_{ij}^*=-e_{ij}^0-m_j^2c_{ij}^1-m_i^2d_{ij}^1-(m_i^2+m_j^2)e_{ij}^1
-m_im_jf_{ij}^1,\nonumber \\
&&f_{ij}^*=-f_{ij}^0-m_im_jc_{ij}^1-m_im_jd_{ij}^1-m_im_je_{ij}^1
-(m_i^2+m_j^2)f_{ij}^1.
\label{solu1}
\end{eqnarray}
From Eq.\ref{counter} and Eq.\ref{solu1}, the renormalized self-energy
is recast into the form
\begin{eqnarray}
&&\Sigma^{REN}_{ij}(k)=\Big(m_i^2d_{ij}^1+m_i^2e_{ij}^1+m_im_jf_{ij}^1
+c_{ij}^1k^2\Big)m_j\omega_- \nonumber \\
&&\hspace{1.5cm}+\Big(m_j^2c_{ij}^1+m_j^2e_{ij}^1+m_im_jf_{ij}^1
+d_{ij}^1k^2\Big)m_i\omega_+ \nonumber \\
&&\hspace{1.5cm}+\Big(-m_j^2c_{ij}^1-m_i^2d_{ij}^1-(m_i^2+m_j^2)e_{ij}^1
-m_im_jf_{ij}^1+e_{ij}^1k^2\Big)/\!\!\!k\omega_- \nonumber \\
&&\hspace{1.5cm}+\Big(-m_im_jc_{ij}^1-m_im_jd_{ij}^1-m_im_je_{ij}^1
-(m_i^2+m_j^2)e_{ij}^1+f_{ij}^1k^2\Big)/\!\!\!k\omega_+\nonumber \\
&&\hspace{1.5cm}=(/\!\!\!k-m_j)\hat{\Sigma}_{ij}(k)(/\!\!\!k-m_i).
\label{sigren}
\end{eqnarray}
In the last step, we have written  $\Sigma_{ij}^{REN}(k)$ in such a way
that their on-shell behavior becomes more obvious as
\begin{eqnarray}
&&\hat{\Sigma}_{ij}(k)=c_{ij}^1m_j\omega_{+}+d_{ij}^1m_i\omega_-
+e_{ij}^1(m_i\omega_-+m_j\omega_++/\!\!\!k\omega_+)\nonumber \\
&&\hspace{1.5cm}+f_{ij}^1(m_i\omega_++m_j\omega_-+/\!\!\!k\omega_-).
\label{hatsig}
\end{eqnarray}
For convenience, we introduce some new symbols:
\begin{eqnarray}
&&\delta Z_{ij}^{L}=-m_j^2c_{ij}^1-m_i^2d_{ij}^1-(m_i^2+m_j^2)e_{ij}^1
-m_im_jf_{ij}^1+e_{ij}^1k^2,\nonumber \\
&&\delta Z_{ij}^R=-m_im_jc_{ij}^1-m_im_jd_{ij}^1-m_im_je_{ij}^1
-(m_i^2+m_j^2)f_{ij}^1+f_{ij}^1k^2,\nonumber \\
&&\delta m_{ij}^{L}=\Big(m_i^2d_{ij}^1+m_i^2e_{ij}^1+m_im_jf_{ij}^1
+c_{ij}^1k^2\Big)m_j,\nonumber \\
&&\delta m_{ij}^R=\Big(m_j^2c_{ij}^1+m_j^2e_{ij}^1+m_im_jf_{ij}^1
+d_{ij}^1k^2\Big)m_i.
\label{symbel}
\end{eqnarray}
Up to one-loop order, the two-point Green's function is
\begin{eqnarray}
&&\Gamma_{ij}(k)=\Big(/\!\!\!k-m_i^{tree}\Big)\delta_{ij}+\Sigma_{ij}^{REN}(k)
\nonumber \\
&&\hspace{1.0cm}=\Big(/\!\!\!k-m_i^{tree}\Big)\delta_{ij}+\delta Z_{ij}^{L}
/\!\!\!k\omega_-+\delta Z_{ij}^R/\!\!\!k\omega_+
-\delta m_{ij}^L\omega_--\delta m_{ij}^R\omega_+
\nonumber \\
&&\hspace{1.0cm}=(\delta_{ij}+\delta Z_{ij}^L)\Big(/\!\!\!k-m_i^{tree}
-\delta m_{ij}^L+\delta Z_{ij}^Lm_i^{tree}\Big)\omega_- \nonumber \\
&&\hspace{1.5cm}+(\delta_{ij}+\delta Z_{ij}^R)\Big(/\!\!\!k-m_i^{tree}
-\delta m_{ij}^R+\delta Z_{ij}^Rm_j^{tree}\Big)\omega_+,
\label{twopoint}
\end{eqnarray}
where $\delta_{ij}+\delta Z_{ij}^{L}$ is the renormalization
multiplier
for the left-handed wave function and $\delta_{ij}+\delta Z_{ij}^{R}$ is
the renormalization multiplier for the right-handed wave function. $m_i^{tree}$
is the mass of the i-th generation of fermions
at tree level. From Eq.\ref{twopoint} and
the mass-shell conditions, we derive the loop corrections to the mass matrix
elements as:
\begin{eqnarray}
&&\delta m_{ij}^{loop}=\bigg\{\Big[\delta m_{ij}^L
+\delta m_{ij}^R\Big]_{k^2=0}-\Big[m_i
\delta Z_{ij}^L|_{k^2=m_i^2}+m_j\delta Z_{ij}^R|_{k^2=m_j^2}
\Big]\bigg\}\nonumber \\
&&\hspace{1.cm}=3m_i^{tree}(m_j^{tree})^2c_{ij}^1+(m_i^{tree}m_j^{tree}
+(m_i^{tree})^2+(m_j^{tree})^2)m_i^{tree}d_{ij}^1
\nonumber \\
&&\hspace{1.5cm}+((m_i^{tree})^2m_j^{tree}+3m_i^{tree}(m_j^{tree})^2)e_{ij}^1
+(3(m_i^{tree})^2m_j^{tree}+m_i^{tree}(m_j^{tree})^2)f_{ij}^1,
\label{loopmass}
\end{eqnarray}
in which $\delta Z_{ij}^{L,R}$, $\delta m_{ij}^{L,R}$ are defined in
Eq.\ref{symbel}. Note that the above formulae
are correct for any type of fermions.
Eq.\ref{loopmass} is the key formulation to compute
the one-loop corrections for the mass matrix of neutrino-neutralino.

In order to compute the form factors $c_{ij}^{0,1}$, $d_{ij}^{0,1}$,
$e_{ij}^{0,1}$ and $f_{ij}^{0,1}$, the one-loop self energy
diagrams should be precisely calculated. The exchanged bosons
in the diagrams can be
either vector or scalar and they correspond to different integrals.
The integral for exchanging vector-boson is
\begin{eqnarray}
&&Amp_{v}(k)=(\mu_{w})^{2\epsilon}\int\frac{d^DQ}{(2\pi)^D}(iA_{\sigma_1}^{(
{\tiny V})}\gamma_{\mu}\omega_{\sigma_1})\frac{i(/\!\!\!Q+/\!\!\!k+m_f)}
{(Q+k)^2-m_f^2}(iB_{\sigma_2}^{({\tiny V})}\gamma^{\mu}\omega_{\sigma_2})
\frac{-i}{Q^2-m_{{\tiny V}}^2}\nonumber \\
&&\hspace{1.5cm}=-\int_0^1dx\int\frac{d^DQ}{(2\pi)^D}\frac{1}{(Q^2+x(1-x)k^2
-xm_f^2-(1-x)m_{\tiny V}^2)^2}\Big\{(2-D)A_{\sigma}^{({\tiny V})}
B_{\sigma}^{({\tiny V})}(1-x)/\!\!\!k\omega_{\sigma}
\nonumber \\
&&\hspace{2.0cm}+Dm_f
A_{\bar{\sigma}}^{({\tiny V})}B_{\sigma}^{({\tiny V})}\omega_{\sigma}\Big\}
\nonumber \\
&&\hspace{1.5cm}=-i\int_0^1dx\int\frac{d^DQ}{(2\pi)^D}\frac{1}{(Q^2
+xm_f^2+(1-x)m_{\tiny V}^2)^2}\Big\{1+\frac{2x(1-x)k^2}{Q^2
+xm_f^2+(1-x)m_{\tiny V}^2}\Big\}\nonumber \\
&&\hspace{2.0cm}\Big\{(2-D)A_{\sigma}^{({\tiny V})}
B_{\sigma}^{({\tiny V})}(1-x)/\!\!\!k\omega_{\sigma}+Dm_f
A_{\bar{\sigma}}^{({\tiny V})}B_{\sigma}^{({\tiny V})}\omega_{\sigma}\Big\}
\label{ampv1}
\end{eqnarray}
where  $D=4-2\epsilon$ and $\mu_w$ represents the
renormalization scale. $A_{\sigma}^{({\tiny V})},\; B_{\sigma}^{
({\tiny V})}$ with $\sigma=\pm$ are the interaction vertices.  $m_{\tiny V}$
represents the mass of the vector boson that appears in the loop and
$m_f$ is for the fermion in the loop. From Eq.\ref{form}, Eq.\ref{series}
and Eq.\ref{ampv1}, we get
\begin{eqnarray}
&&c_{ij}^0(m_{\tiny V},m_f)=-iD\frac{m_f}{m_j}A_{+}^{({\tiny V})}
B_{-}^{({\tiny V})}F_{2a}(m_f,m_{\tiny V}),\nonumber \\
&&d_{ij}^0(m_{\tiny V},m_f)=-iD\frac{m_f}{m_i}A_{-}^{({\tiny V})}
B_{+}^{({\tiny V})}F_{2a}(m_f,m_{\tiny V}),\nonumber \\
&&e_{ij}^0(m_{\tiny V},m_f)=-i(2-D)A_{-}^{({\tiny V})}
B_{-}^{({\tiny V})}F_{2b}(m_f,m_{\tiny V}),\nonumber \\
&&f_{ij}^0(m_{\tiny V},m_f)=-i(2-D)A_{+}^{({\tiny V})}
B_{+}^{({\tiny V})}F_{2b}(m_f,m_{\tiny V}),\nonumber \\
&&c_{ij}^1(m_{\tiny V},m_f)=-i4\frac{m_f}{m_j}A_{+}^{({\tiny V})}
B_{-}^{({\tiny V})}F_{3a}(m_f,m_{\tiny V}),\nonumber \\
&&d_{ij}^1(m_{\tiny V},m_f)=-i4\frac{m_f}{m_i}A_{-}^{({\tiny V})}
B_{+}^{({\tiny V})}F_{3a}(m_f,m_{\tiny V}),\nonumber \\
&&e_{ij}^1(m_{\tiny V},m_f)=i2A_{-}^{({\tiny V})}
B_{-}^{({\tiny V})}F_{3b}(m_f,m_{\tiny V}),\nonumber \\
&&f_{ij}^1(m_{\tiny V},m_f)=i2A_{+}^{({\tiny V})}
B_{+}^{({\tiny V})}F_{3b}(m_f,m_{\tiny V}).
\label{form1}
\end{eqnarray}
$F_{2a},F_{2b},F_{3a}$ and $F_{3b}$ are integrals over the
internal momentum of the loop and their explicit forms are given
in appendix A.

For exchanging scalar-boson, the amplitude is derived
in a similar way and it is
\begin{eqnarray}
&&Amp_{s}(k)=(\mu_{w})^{2\epsilon}\int\frac{d^DQ}{(2\pi)^D}(iA_{\sigma_1}^{(
{\tiny S})}\omega_{\sigma_1})\frac{i(/\!\!\!Q+/\!\!\!k+m_f)}
{(Q+k)^2-m_f^2}(iB_{\sigma_2}^{({\tiny S})}\omega_{\sigma_2})
\frac{i}{Q^2-m_{{\tiny S}}^2}\nonumber \\
&&\hspace{1.5cm}=i\int_0^1dx\int\frac{d^DQ}{(2\pi)^D}\frac{1}{(Q^2
+xm_f^2+(1-x)m_{\tiny S}^2)^2}\Big\{1+\frac{2x(1-x)k^2}{Q^2
+xm_f^2+(1-x)m_{\tiny S}^2}\Big\}\nonumber \\
&&\hspace{2.0cm}\Big\{A_{\bar{\sigma}}^{({\tiny S})}
B_{\sigma}^{({\tiny S})}(1-x)/\!\!\!k\omega_{\sigma}+m_f
A_{\sigma}^{({\tiny S})}B_{\sigma}^{({\tiny
S})}\omega_{\sigma}\Big\},
\label{amps1}
\end{eqnarray}
where $A_{\sigma}^{({\tiny S})},\; B_{\sigma}^{({\tiny S})}$ with
$\sigma=\pm$ are the interaction vertices.  $m_{\tiny S}$
represents the mass of the scalar boson that appears in the loop and
$m_f$ is for the fermion in the loop.
From Eq.\ref{form}, Eq.\ref{series} and Eq.\ref{amps1}, we
obtain
\begin{eqnarray}
&&c_{ij}^0(m_{\tiny S},m_f)=i\frac{m_f}{m_j}A_{-}^{({\tiny S})}
B_{-}^{({\tiny S})}F_{2a}(m_f,m_{\tiny S}),\nonumber \\
&&d_{ij}^0(m_{\tiny S},m_f)=i\frac{m_f}{m_i}A_{+}^{({\tiny S})}
B_{+}^{({\tiny S})}F_{2a}(m_f,m_{\tiny S}),\nonumber \\
&&e_{ij}^0(m_{\tiny S},m_f)=iA_{+}^{({\tiny S})}
B_{-}^{({\tiny S})}F_{2b}(m_f,m_{\tiny S}),\nonumber \\
&&f_{ij}^0(m_{\tiny V},m_f)=iA_{-}^{({\tiny S})}
B_{+}^{({\tiny S})}F_{2b}(m_f,m_{\tiny S}),\nonumber \\
&&c_{ij}^1(m_{\tiny S},m_f)=i\frac{m_f}{m_j}A_{-}^{({\tiny S})}
B_{-}^{({\tiny S})}F_{3a}(m_f,m_{\tiny S}),\nonumber \\
&&d_{ij}^1(m_{\tiny S},m_f)=i\frac{m_f}{m_i}A_{+}^{({\tiny S})}
B_{+}^{({\tiny S})}F_{3a}(m_f,m_{\tiny S}),\nonumber \\
&&e_{ij}^1(m_{\tiny S},m_f)=iA_{+}^{({\tiny S})}
B_{-}^{({\tiny S})}F_{3b}(m_f,m_{\tiny S}),\nonumber \\
&&f_{ij}^1(m_{\tiny V},m_f)=iA_{-}^{({\tiny S})}
B_{+}^{({\tiny S})}F_{3b}(m_f,m_{\tiny S}).
\label{form2}
\end{eqnarray}
In the supersymmetric extension of the SM, the mixing of $\kappa_i^0
\sim\kappa_j^0$ originates from the following loop-diagrams.

\begin{itemize}
\item The internal particles are $Z\sim$gauge boson and neutralinos
(or neutrinos through mixing) $\kappa_\alpha^0$ ($\alpha=1,2,\cdots,7$).
\item  The internal particles are $W\sim$gauge boson and charginos
(or charged leptons) $\kappa_\alpha^0$ ($\alpha=1,2,\cdots,5$).
\item The internal particles are CP-even Higgs bosons $H_\beta^0$ ($\beta=1,2,
\cdots,5$) and neutralinos (or neutrinos through mixing) $\kappa_\alpha^0$
($\alpha=1,2,\cdots,7$).
\item The internal particles are CP-odd Higgs bosons $A_\beta^0$ ($\beta=1,2,
\cdots,5$) and neutralinos (or neutrinos through mixing) $\kappa_\alpha^0$
($\alpha=1,2,\cdots,7$).
\item The internal particles are charged Higgs bosons $H_\beta^+$ ($\beta=1,2,
\cdots,8$) and charginos (or charged leptons) $\kappa_\alpha^-$
($\alpha=1,2,\cdots,5$).
\item The internal particles are up-type scalar quarks $\tilde{U}^k_\alpha$
($\alpha=1,2$; k=1, 2, 3.)  and quark $u^k$.
\item The internal particles are down-type scalar quarks $\tilde{D}^k_\alpha$
($\alpha=1,2$; k=1, 2, 3.)  and quark $d^k$.
\end{itemize}
The expressions of the contributions from
those loop diagrams to the self-energy are presented in Appendix B.

Summing over Eq.\ref{nzcor},  Eq.\ref{cwcor}, Eq.\ref{cpencor}, Eq.\ref{cponcor},
Eq.\ref{chccor}, Eq.\ref{upqcor} and Eq.\ref{dwqcor} in Appendix
B, we  obtain the one-loop corrections
to the neutrino-neutralino mass matrix
\begin{equation}
\delta m_{ij}^{1-loop}=\delta m_{ij}^{(Z,\kappa^0)}+\delta m_{ij}^{(W,\kappa^-)}
+\delta m_{ij}^{(H^0,\kappa^0)}+\delta m_{ij}^{(A^0,\kappa^0)}+
\delta m_{ij}^{(H^+,\kappa^-)}+\delta m_{ij}^{(\tilde{U},u)}
+\delta m_{ij}^{(\tilde{D},d)}.
\label{sum}
\end{equation}
From the above analysis, we find that
the one-loop corrections to the neutrino-neutralino mass matrix are
decreasing when  $m_S$ (S represents the scalar SUSY or Higgs
particles) turns heavier.
As a pioneer work, the authors of \cite{add} discussed the
neutrino problems in the supersymmetric model without R-parity.
Even though our method is different from theirs,  our
results are qualitatively consistent with theirs, namely,
as the mass of the SUSY particles increases,
the mass of neutrinos decreases.

\section{The masses of  neutrinos\label{model}}
In this section, we  discuss the diagonalization of the mass matrix
of neutrino-neutralino with one-loop correction. For illustrating the physics
picture, we
first consider a simplified model where only $\tau$-number is violated, and
then generalize the result to the case of three-generation lepton
number violation.
\subsection{A simplified model: only the $\tau$-number is violated}

In the mass basis, the one-loop correction to the neutrino-neutralino
can be written as
\begin{equation}
{\cal M}_{N}^{1-loop}=\left(
\begin{array}{ccccc}
m_{\kappa^0_1}^{tree} & 0 & 0 & 0 & \delta m_{15}\\
0 & m_{\kappa^0_2}^{tree} & 0 & 0 & \delta m_{25}\\
0 & 0 &m_{\kappa^0_3}^{tree}& 0 & \delta m_{35}\\
0 & 0 & 0 & m_{\kappa^0_4}^{tree}& \delta m_{45}\\
\delta m_{15} & \delta m_{25} & \delta m_{35}
& \delta m_{45} &m_{\nu_{\tau}}^{tree}+\delta m_{55}\\
\end{array}
\right).
\label{toy1}
\end{equation}
In fact, for neutralinos are much heavier than neutrinos and the
loop-corrections to them are very small, so that
here, we have ignored the loop corrections to the masses
of neutralinos. From Eq.\ref{toy1}, we can get
\begin{equation}
m_{\nu_{\tau}}^{1-loop}=m_{\nu_{\tau}}^{tree}+\delta m_{55}
-\frac{\delta m_{15}^2}{m_{\kappa_1^0}}
-\frac{\delta m_{25}^2}{m_{\kappa_2^0}}
-\frac{\delta m_{35}^2}{m_{\kappa_3^0}}
-\frac{\delta m_{45}^2}{m_{\kappa_4^0}}.
\end{equation}
All the off-diagonal mass terms and extra diagonal masses  $\delta m_{ij}$
are given in Eq.(\ref{sum}). Different from the results in the
$\overline{MS}$-scheme and the approach with the mass-insertion method, the
on-shell scheme may give rise to larger corrections to the
off-diagonal matrix elements $\delta m_{ij}(i\neq j)$ than to the
diagonal elements for small neutrino masses. We will give more
discussions on the difference in the last section.
The corrections from
W,Z-gauge bosons and Goldstone particles are suppressed by the
small mixing  between the neutrino-neutralino. Now, we discuss the
scalar particle contributions. For convenience, we restrain our
discussion in the basis where $\upsilon_L=0$ (the vacuum
expectation value of the scalar lepton). When the CP-even Higgs
mass $m_{H_\beta^0}\gg m_{\kappa_\alpha^0}$, the off-diagonal
correction from $H_{\beta}^0$ is
\begin{eqnarray}
&&\delta m_{i5}\sim \frac{1}{(4\pi)^2}\frac{m_{\kappa_i^0}^2}{2m_{H_\beta^0}^2}
(c_{w}Z_N^{j2}-s_wZ_N^{j1})\Big(\sum\limits_{\delta=1}^5Z_E^{\beta \delta}
Z_N^{i(2+\delta)}-2Z_E^{\beta 2}Z_N^{i3}\Big)\Big[c_wZ_E^{\beta 2}
\frac{1}{2}g\upsilon_1\nonumber \\
&&\hspace{1.5cm}+c_wZ_E^{\beta 1}\frac{1}{2}g\upsilon_2
+s_wZ_E^{\beta 2}
\frac{1}{2}g'\upsilon_1+s_wZ_E^{\beta 1}\frac{1}{2}g'\upsilon_2
\Big]\frac{e^2}{4s_w^2c_w^2}\nonumber \\
&&\hspace{1.cm}\sim
\frac{\alpha_e}{32\pi}\frac{m_{\kappa_i^0}^2m_w}{m_{H_\beta^0}^2}.
\label{cehig1}
\end{eqnarray}
As $\frac{\delta m_{i5}^2}{m_i}\sim 10^{-9}$(GeV) is required,
we have $m_{H_\beta^0}\geq 3m_i$.
The corrections from the physical CP-odd Higgs bosons, Charged Higgs bosons
can be discussed in a similar way, and they are suppressed by the large
masses of the scalar bosons. For small neutrino masses,
one could have another solution where all the R-parity violation parameters (in
soft breaking terms and the superpotential) are all small, thus their
contributions can be neglected, this scenario has been discussed by
Davidson {\it et.al} \cite{Davidson}.
\subsection{The model with three-generation lepton number violation}

In this case, the mass matrix of neutrino-neutralino can be written as
\begin{equation}
{\cal M}_{N}^{1-loop}=\left(
\begin{array}{ccccccc}
m_{\kappa^0_1}^{tree} & 0 & 0 & 0 & \delta m_{15}&\delta m_{16}&\delta m_{17}\\
0&m_{\kappa^0_2}^{tree} & 0 & 0 & \delta m_{25}&\delta m_{26}&\delta m_{27}\\
0&0&m_{\kappa^0_3}^{tree}& 0 & \delta m_{35}&\delta m_{36}&\delta m_{37}\\
0&0&0&m_{\kappa^0_4}^{tree} &\delta m_{45}&\delta m_{46}&\delta m_{47}\\
\delta m_{15}&\delta m_{25}&\delta m_{35}&\delta m_{45}
&m_{\nu_{\tau}}^{tree}+\delta m_{55}&\delta m_{56}&\delta m_{57}\\
\delta m_{16}&\delta m_{26}&\delta m_{36}&\delta m_{46}
&\delta m_{56}&0&0\\
\delta m_{17}&\delta m_{27}&\delta m_{37}&\delta m_{47}
&\delta m_{57}&0&0\\
\end{array}
\right).
\label{nmass}
\end{equation}
The eigenequation of the matrix (\ref{nmass}) is
\begin{eqnarray}
&&x^3-\Big(m_{\nu_{\tau}}^{tree}+\delta m_{55}-\sum\limits_{i=1}^4
\frac{\delta m_{i5}^2+\delta m_{i6}^2+\delta m_{i7}^2}{m_{\kappa_i^0}^{tree}}
\Big)x^2\nonumber \\
&&-\Big(\sum\limits_{i=1}^4\frac{(\delta m_{i6}^2+\delta m_{i7}^2)
(m_{\nu_{\tau}}^{tree}+\delta m_{55})-2\delta m_{i5}\delta m_{i6}
\delta m_{56}-2\delta m_{i5}\delta m_{i7}\delta m_{57}}{m_{\kappa_i^0}^{tree}}
\nonumber \\
&&+(\delta m_{56}^2+\delta m_{57}^2)\Big)x-\sum\limits_{i=1}^4
\frac{(\delta m_{i7}\delta m_{56}-\delta m_{i6}\delta m_{57})^2}
{m_{\kappa_i^0}^{tree}}=0.
\label{eigen}
\end{eqnarray}
It is noted that even though $M_N^{1-loop}$ is a
7th-order matrix, because of  the
form of the up-left corner submatrix, this equation can be reduced
into a 3rd-order one.

In a general theory, one can notice from the analysis given above that
the lepton-number violation  terms (bilinear or trilinear)
and the corresponding soft-violating terms affect
the mass matrix of neutrinos through mixings of
neutrino-neutralino, charged lepton-chargino and Higgs-slepton.
However, in this general scenario, there are too many parameters
to reduce the prediction power of the model. Thus in our later
analysis, we ignore the trilinear terms
$\lambda^{ikl}\hat L_i\hat L_k\hat E^c_l$ ($i,k,l=1,2,3$),
$\lambda^{jpq}_d\hat L_j\hat Q_p\hat D_q^c$ ($j,p,q=1,2,3$)
in the superpotential and the
corresponding soft-violating terms.

In terms of the mass insertion method, the authors of \cite{Haug}
discussed the modification to the neutrino mass
caused by influence of the R-parity
violating trilinear terms of the superpotential in every details.
For example, if only the third generation of the down-type scalar quarks is
relatively light, the modification of the $q\tilde q-$loops
induced by  the R-parity violating
trilinear terms and the corresponding soft-violating terms of the
superpotential to the neutrino mass matrix is
\begin{equation}
\delta m^{q\tilde q}_{(8-i)(8-j)}\sim
{3\lambda_d^{i33}\lambda_d^{j33}\over 8\pi^2} {m_b^2(A_d^3+
\mu^{0}\tan\beta)\over m_{\tilde b}^2}.
\end{equation}
Similarly, if the third generation of charged sleptons is the
lightest, the contribution of $l\tilde l-$loops induced by
$\lambda^{ijk}\hat L_i\hat L_j\hat E_k^c$ and the corresponding
soft-violating terms is
\begin{equation}
\delta m^{l\tilde l}_{(8-i)(8-j)}\sim
{3\lambda^{i33}\lambda^{j33}\over 8\pi^2} {m_{\tau}^2(A_l^3+
\mu^{0}\tan\beta)\over m_{\tilde{\tau}}^2}.
\end{equation}
Because of the method and renormalization scheme adopted in our
calculations, our results are a bit different from that obtained
in terms of the mass-insertion method. As a matter of fact, in our
calculations, if the R-parity violating trilinear terms are also
taken into account, their effects change the mass matrix of
neutrino through their modifications to the
effective interaction vertices and thus, as a reasonable approximation,
can be included in the
effective couplings of trilinear interaction vertices.

Now let us turn to the numerical computations.
We diagonalize the mixing matrix
and obtain the neutrino mass eigenvalues of three
generations as well as the modified
neutralino masses. This is a generalized see-saw mechanism.

Numerically, we adopt the basis with $\upsilon_{\tilde{\nu}_i}=0.$
({\it i}=1, 2, 3) and input  the soft-breaking terms as
$$m_{L}^2(0,0)=-6.0\times 10^6\;{\rm GeV^2},\;\;\;
m_{L}^2(1,1)=6.0\times 10^6\;{\rm GeV^2},$$
$$m_{L}^2(2,2)=6.0\times 10^6\;{\rm GeV^2},\;\;\;
m_{L}^2(3,3)=6.0\times 10^6\;{\rm GeV^2},$$
$$m_{L}^2(0,1)=0.\;{\rm GeV^2},\;\;\; m_{L}^2(0,2)=0.\;{\rm GeV^2},\;\;\;
m_{L}^2(0,3)=2.0\times 10^5\;{\rm GeV^2},$$
$$m_{L}^2(1,2)=0.\;{\rm GeV^2},\;\;\; m_{L}^2(1,3)=0.\;{\rm GeV^2},\;\;\;
m_{L}^2(2,3)=0.\;{\rm GeV^2}.$$

We also set the R-parity conserving term $$ B^0=10^6\;{\rm
GeV^2},$$ and the R-parity violation parameters $B^i$ for
three generations in the super-potential as
$$B^1=B^2=10\;{\rm GeV^2}\;\;\;,$$
and $B^3$ is treated as a variable which can be fixed by comparing
the calculated results with data.
The choice of the soft
breaking parameters in our scenario makes the lepton-number
violation effects of the 1st and 2nd generation fermions
are much less significant than that of the 3rd generation
\cite{diaz}.

Several other concerned parameters are taken as
$$\tan\beta=5,\;\;\mu^0=100\; {\rm
GeV},\;\;\mu^i=0\;\;(i=1,2),\;\; \mu^3=10^{-6}\;{\rm
GeV},$$ This choice leads to  a nonzero mass less than 1 eV
for  the 3rd generation neutrino at the tree order.

Then we have obtained
$$m_{\chi_1^0}^{tree}\simeq 44.0\;{\rm GeV},$$
$$m_{\chi_2^0}^{tree}\simeq 99.0\;{\rm GeV},$$
$$m_{\chi_3^0}^{tree}\simeq 117.0\;{\rm GeV},$$
$$m_{\chi_4^0}^{tree}\simeq 172.0\;{\rm GeV}.$$

The corresponding dependence of $\Delta m^2_{23}\equiv
|m^2_{\nu_{\tau}}-m^2_{\nu_{\mu}}|$ and $\Delta
m^2_{12}\equiv |m^2_{\nu_{\mu}}-m^2_{\nu_{e}}|$ on the
R-parity violation parameter of the third generation $B^3$
is shown in Fig.1. Simultaneously, we can have the mixing
angles as $$|\sin\theta_{23}|\simeq
0.26,\;\;\;\;|\sin\theta_{12}|\simeq 0.034.$$ In the
numerical analyses, we find that the mixing angles
$\theta_{23}$, $\theta_{12}$ do not vary much as the soft
breaking parameter $B_3$ changes, and this small declination
can be neglected in the physical discussions.

Therefore we can draw a rough conclusion that as the
R-parity violation parameter of the third generation $B^3$
is larger than $10^{5}$ GeV$^2$, the obtained neutrino mass
difference $\Delta m^2_{23}$ and mixing angles can meet the
data. From Fig.1, we  obtain that when $B^3>10^5$ GeV,
the neutrino mass differences
 $$\Delta m^2_{23}\sim 10^{-1}\;{\rm eV^2}\;\;\;
{\rm and}\;\;\; \Delta m^2_{12}\sim 10^{-6}\;{\rm eV^2}.$$
The recent data on the solar neutrino and atmospheric
neutrino experiments have been given in \cite{Mohapatra}. The
more precise values of the oscillation parameters at 90\%
c.l. are:
\begin{eqnarray}
\Delta m^2_{\nu_{\mu}\nu_{\tau}}\simeq (2~-~8)\times 10^{-3}~ eV^2;\\
\nonumber
sin^2 2\theta_{\mu\tau}\simeq 0.8~-~1
\end{eqnarray}
and the best fit to data for the solar neutrino experiments
gives several possible solutions
corresponding to various models as
\begin{eqnarray}
&&VO:  \Delta m^2 \simeq 6.5\times 10^{-11} eV^2; sin^22\theta\simeq 0.75
-1 \\ \nonumber
&&SMA-MSW: \Delta m^2 \simeq 5\times 10^{-6} eV^2;  sin^22\theta \simeq
5\times 10^{-3} \\ \nonumber
&&LMA-MSW:  \Delta m^2 \simeq 1.2\times 10^{-5}-3.1\times 10^{-4} eV^2;
sin^22\theta \simeq 0.58~-~1.00,
\end{eqnarray}
where "VO" means "vacuum oscillation" solution, "SMA-MSW"
denotes the solution of Mikheev-Smirnov-Wolfenstein
with small mixing angle whereas "LMA-MSW" is similar to
"SMA-MSW" but with larger mixing angle.

Our results for $\Delta m^2_{23}$ is about $10^{-2}$ eV$^2$ for
$B^3>3\times 10^5$ GeV$^2$ and $\Delta m^2_{12}$ is $10^{-5}$ eV$^2$
at the scale. The mixing angle is  $\sin^22\theta_{23}\sim 0.27$ which is
a bit smaller than the fitted data, but has the same order of magnitude.
Due to the relatively larger errors in both experiments and
theory, this difference is tolerable.
The results of $\Delta m^2_{12}$, $\sin^22\theta_{21}
\sim 4.6\times 10^{-3}$ in our calculation are consistent
with the case of SMA-MSW very well.

\section{Conclusion and discussion}

In this work, we employ the on-shell renormalization scheme.
We find that the supersymmetric model with L-number violation
terms can result in the expected values which are gained by
fitting the solar and atmospheric neutrino experiments.
Even though the results are qualitatively consistent with that in
the $\overline{MS}$ scheme, quantitatively  they are a bit different
from the later. To our understanding it is because the
on-shell renormalization scheme is an over-subtraction scheme,
after eliminating the divergence of the loop integrations, at any
concerned order of perturbation (the one-loop order in our case),
the equation of motion of the particle still holds, i.e. the
mass-shell condition is respected. Instead, in the $\overline{MS}$
scheme, the mass-shell condition is violated at the concerned order
of perturbation, and this may be the reason of the difference
in the two schemes.

When we take into account the contribution from the R-parity violating
trilinear terms in our numerical analysis, the available parameter
space would be enlarged, it is easy to find a suitable subspace of
concerned parameters which can fit all present experimental data.
In our future studies, we may extend our analysis to other
phenomenology where the R-parity violating terms play important
roles and then, maybe, the available parameter space is
more constrained
and  existence of both the bilinear and trilinear terms would be
necessary. By comparison, in this work, their effects merge into the effective
couplings of the vertices for the bilinear interactions, so that
do not manifest themselves directly.

More studies on this subject need to be carried out,
especially the parameter space for SUSY without R-parity is
tremendously large, thus more precise measurements on neutrino
oscillation as well as other rare B-decays
 may help to understand both neutrino physics and SUSY
without R-parity.

\vspace{1.0cm}
\noindent {\Large{\bf Acknowledgements}}

This work is partially supported by the National Natural Science
Foundation of China.

\vspace{1cm}

\appendix
{\Large{\bf Appendix}}

\section{The definition of $F_{2a}$, $F_{2b}$, $F_{3a}$, $F_{3b}$}

The functions that appear in the text are defined as
\begin{eqnarray}
&&F_{2a}(m_1,m_2)=(\mu_w)^{2\epsilon}\int_0^1dx\int\frac{d^DQ}{(2\pi)^D}
\frac{1}{(Q^2+xm_1^2+(1-x)m_2^2)^2}\nonumber \\
&&\hspace{2.0cm}=\frac{1}{(4\pi)^2}\Big\{\frac{1}{\epsilon}-\gamma_E
+\frac{m_1^2\ln\frac{4\pi\mu_{w}^2}{m_1^2}-m_2^2\ln\frac{4\pi\mu_{w}^2}
{m_2^2}}{m_1^2-m_2^2}\Big\},\nonumber \\
&&F_{2b}(m_1,m_2)=(\mu_w)^{2\epsilon}\int_0^1dx\int\frac{d^DQ}{(2\pi)^D}
\frac{1-x}{(Q^2+xm_1^2+(1-x)m_2^2)^2}\nonumber \\
&&\hspace{2.0cm}=\frac{1}{2(4\pi)^2}\Big\{\frac{1}{\epsilon}-\gamma_E
+\frac{m_1^2-3m_2^2}{2(m_1^2-m_2^2)}\nonumber \\
&&\hspace{2.5cm}+\frac{1}{(m_1^2-m_2^2)^2}\Big[
m_1^2(m_1^2-2m_2^2)\ln\frac{4\pi\mu_{w}^2}{m_1^2}+m_2^2\ln\frac{4\pi
\mu_{w}^2}{m_2^2}\Big]\Big\},\nonumber \\
&&F_{3a}(m_1,m_2)=(\mu_w)^{2\epsilon}\int_0^1dx\int\frac{d^DQ}{(2\pi)^D}
\frac{2x(1-x)}{(Q^2+xm_1^2+(1-x)m_2^2)^3}\nonumber \\
&&\hspace{2.0cm}=\frac{1}{(4\pi)^2}\frac{1}{(m_1^2-m_2^2)^3}\Big\{
-m_1^2m_2^2\ln\frac{m_1^2}{m_2^2}+\frac{m_1^4-m_2^4}{2}\Big\},
\nonumber \\
&&F_{3b}(m_1,m_2)=(\mu_w)^{2\epsilon}\int_0^1dx\int\frac{d^DQ}{(2\pi)^D}
\frac{2x(1-x)^2}{(Q^2+xm_1^2+(1-x)m_2^2)^3}\nonumber \\
&&\hspace{2.0cm}=\frac{1}{(4\pi)^2}\frac{1}{(m_1^2-m_2^2)^4}\Big\{
\frac{1}{3}m_1^6+\frac{1}{6}m_2^6+\frac{1}{2}m_1^4m_2^2-m_1^2m_2^4
-m_1^4m_2^2\ln\frac{m_1^2}{m_2^2}\Big\}.
\label{function}
\end{eqnarray}

\section{The 1-loop self energy diagrams of neutralino-neutrino
mixing\label{oloop}}

\begin{itemize}
\item The internal particles are $Z\sim$gauge boson and neutralinos
(neutrinos through mixing) $\kappa_\alpha^0$ ($\alpha=1,2,\cdots,7$). The coupling
constants can be written as
\begin{eqnarray}
&&A_-^{({\tiny Z, \kappa_{\alpha}^0})}=\frac{e}{4s_{w}c_{w}}
\bigg\{Z_{N}^{*\alpha,3}Z_{N}^{i,3}-\sum\limits_{\delta=4}^{7}
Z_{N}^{*\alpha,\delta}Z_{N}^{i,\delta}\bigg\},\nonumber \\
&&A_+^{({\tiny Z, \kappa_{\alpha}^0})}=
-A_-^{({\tiny Z, \kappa_{\alpha}^0})},\nonumber \\
&&B_-^{({\tiny Z, \kappa_{\alpha}^0})}=\frac{e}{4s_{w}c_{w}}
\bigg\{Z_{N}^{*j,3}Z_{N}^{\alpha,3}-\sum\limits_{\delta=4}^{7}
Z_{N}^{*j,\delta}Z_{N}^{\alpha,\delta}\bigg\},\nonumber \\
&&B_+^{({\tiny Z, \kappa_{\alpha}^0})}=
-B_-^{({\tiny Z, \kappa_{\alpha}^0})},\nonumber \\
&&m_{f}=m_{\kappa_{\alpha}^0}, \hspace{2.0cm}m_{V}=m_{z}.
\label{nzloop}
\end{eqnarray}
$Z_N$ is the transformation matrix of neutrino-neutralino
from the interaction basis
to the mass basis. Corrections to the mass matrix are:
\begin{eqnarray}
&&\delta m_{ij}^{(Z,\kappa^0)}=\sum\limits_{\alpha=1}^7
\bigg\{\Big(12m_{\kappa_i^0}^{tree}m_{\kappa_j^0}^{tree}
A_+^{({\tiny Z, \kappa_{\alpha}^0})}B_-^{({\tiny Z, \kappa_{\alpha}^0})}
+((m^{tree}_{\kappa_i^0})^2+m_{\kappa_i^0}^{tree}m_{\kappa_j^0}^{tree}\nonumber \\
&&\hspace{2.cm}+(m_{\kappa_0^j}^{tree})^2)
A_-^{({\tiny Z, \kappa_{\alpha}^0})}B_+^{({\tiny Z, \kappa_{\alpha}^0})}
\Big)m_{\kappa_\alpha^0}F_{3a}(m_{\kappa_{\alpha}^0},m_{Z})\nonumber \\
&&\hspace{2.cm}-2\Big[((m^{tree}_{\kappa_i^0})^2m_{\kappa_j^0}^{tree}
+3m_{\kappa_i^0}^{tree}(m^{tree}_{\kappa_j^0})^2)
A_-^{({\tiny Z, \kappa_{\alpha}^0})}B_-^{({\tiny Z, \kappa_{\alpha}^0})}
\nonumber \\
&&\hspace{2.cm}+(m_{\kappa_i^0}^{tree}(m^{tree}_{\kappa_j^0})^2+3(m^{tree}_{\kappa_i^0})^2
m_{\kappa_j^0}^{tree})A_+^{({\tiny Z, \kappa_{\alpha}^0})}
B_+^{({\tiny Z, \kappa_{\alpha}^0})}
\Big]F_{3b}(m_{\kappa_{\alpha}^0},m_{Z})\Big]
\label{nzcor}
\end{eqnarray}
\item  The internal particles are $W\sim$gauge boson and charginos
(charged leptons) $\kappa_\alpha^0$ ($\alpha=1,2,\cdots,5$). The coupling
constants can be written as
\begin{eqnarray}
&&A_-^{({\tiny W, \kappa_{\alpha}^-})}=\frac{e}{s_{w}}\Big\{
-Z_+^{*\alpha,1}Z_N^{i,2}+\frac{1}{\sqrt{2}}Z_+^{*\alpha,2}Z_N^{i,3}\Big\}
\nonumber \\
&&A_+^{({\tiny W, \kappa_{\alpha}^-})}=\frac{e}{s_{w}}\Big\{
Z_-^{\alpha,1}Z_N^{*i,2}+\frac{1}{\sqrt{2}}\sum\limits_{\delta=0}^3
Z_-^{\alpha,2+\delta}Z_N^{*i,4+\delta}\Big\}
\nonumber \\
&&B_-^{({\tiny W, \kappa_{\alpha}^-})}=\frac{e}{s_{w}}\Big\{
-Z_+^{\alpha,1}Z_N^{*j,2}+\frac{1}{\sqrt{2}}Z_+^{\alpha,2}Z_N^{*j,3}\Big\}
\nonumber \\
&&B_+^{({\tiny W, \kappa_{\alpha}^-})}=\frac{e}{s_{w}}\Big\{
Z_-^{*\alpha,1}Z_N^{j,2}+\frac{1}{\sqrt{2}}\sum\limits_{\delta=0}^3
Z_-^{*\alpha,2+\delta}Z_N^{j,4+\delta}\Big\}\nonumber\\
&&m_f=m_{\kappa_{\alpha}^-},\hspace{2.0cm}m_{\tiny V}=m_{w}
\label{cwloop}
\end{eqnarray}
where  $Z_\pm$ is the transformation matrix of charged lepton-chargino
from the interaction basis to the mass basis.
The W-chargino loop corrections to the mass matrix are:
\begin{eqnarray}
&&\delta m_{ij}^{(W,\kappa^0)}=\sum\limits_{\alpha=1}^5
\bigg\{\Big[12m_{\kappa_i^0}^{tree}m_{\kappa_j^0}^{tree}
A_+^{({\tiny W, \kappa_{\alpha}^-})}B_-^{({\tiny W, \kappa_{\alpha}^-})}
+((m^{tree}_{\kappa_i^0})^2+(m^{tree}_{\kappa_j^0})^2\nonumber \\
&&\hspace{2.cm}+m_{\kappa_i^0}^{tree}m_{\kappa_j^0}^{tree})
A_-^{({\tiny W, \kappa_{\alpha}^-})}B_+^{({\tiny W, \kappa_{\alpha}^-})}
\Big]m_{\kappa_\alpha^-}F_{3a}(m_{\kappa_{\alpha}^-},m_{W})\nonumber \\
&&\hspace{2.cm}+\Big[((m^{tree}_{\kappa_i^0})^2m_{\kappa_j^0}^{tree}+3m_{\kappa_i^0}^{tree}
(m^{tree}_{\kappa_j^0})^2)
A_-^{({\tiny W, \kappa_{\alpha}^-})}B_-^{({\tiny W, \kappa_{\alpha}^-})}
\nonumber \\
&&\hspace{2.cm}
+(3(m^{tree}_{\kappa_i^0})^2m_{\kappa_j^0}^{tree}+m_{\kappa_i^0}^{tree}(m^{tree}_{\kappa_j^0})^2)
A_+^{({\tiny W, \kappa_{\alpha}^-})}
B_+^{({\tiny W, \kappa_{\alpha}^-})}
\Big]F_{3b}(m_{\kappa_{\alpha}^-},m_{W})\Big\}
\label{cwcor}
\end{eqnarray}
\item The internal particles are CP-even Higgs bosons $H_\beta^0$ ($\beta=1,2,
\cdots,5$) and neutralinos (neutrinos through mixing) $\kappa_\alpha^0$
($\alpha=1,2,\cdots,7$). The coupling constants can be written as
\begin{eqnarray}
&&A_-^{({\tiny H_\beta^0,\kappa_\alpha^0})}=\frac{e}{2s_wc_w}
C_{snn}^{\alpha j\beta}\nonumber \\
&&A_+^{({\tiny H_\beta^0,\kappa_\alpha^0})}=\frac{e}{2s_wc_w}
C_{snn}^{*\alpha j\beta}\nonumber \\
&&B_-^{({\tiny H_\beta^0,\kappa_\alpha^0})}=\frac{e}{2s_wc_w}
C_{snn}^{\alpha \beta i}\nonumber \\
&&B_+^{({\tiny H_\beta^0,\kappa_\alpha^0})}=\frac{e}{2s_wc_w}
C_{snn}^{*\alpha \beta i}\nonumber \\
&&m_{\tiny S}=m_{H_\beta^0},\hspace{2.0cm}m_f=m_{\kappa_\alpha^0}
\label{cpenloop}
\end{eqnarray}
with
\begin{eqnarray}
&&C_{snn}^{\alpha\beta\gamma}=\Big(c_wZ_{N}^{\beta 2}-s_wZ_N^{\beta 1}
\Big)\Big(\sum\limits_{\delta=1}^{5}Z_{E}^{\alpha\delta}Z_N^{\gamma 2+\delta}
-2Z_{E}^{\alpha 2}Z_N^{\gamma 3}\Big)
\label{csnn}
\end{eqnarray}
and $Z_E$ is the mixing matrix of CP-even Higgs bosons.
The CP-even Higgs-neutralino loop corrections to the mass matrix
are
\begin{eqnarray}
&&\delta m_{ij}^{(H^0,\kappa^0)}=-\sum\limits_{\alpha=1}^7
\sum\limits_{\beta=1}^5
\bigg\{\Big(3m_{\kappa_i^0}^{tree}m_{\kappa_j^0}^{tree}
A_-^{({\tiny H_{\beta}^0, \kappa_{\alpha}^0})}
B_-^{({\tiny H_{\beta}^0, \kappa_{\alpha}^0})}
+((m^{tree}_{\kappa_i^0})^2+(m^{tree}_{\kappa_j^0})^2\nonumber \\
&&\hspace{2.cm}+m_{\kappa_i^0}^{tree}m_{\kappa_j^0}^{tree})
A_+^{({\tiny H_{\beta}^0, \kappa_{\alpha}^0})}
B_+^{({\tiny H_{\beta}^0, \kappa_{\alpha}^0})}
\Big)m_{\kappa_\alpha^-}F_{3a}(m_{\kappa_{\alpha}^0},m_{H_{\beta}^0})
\nonumber \\
&&\hspace{2.cm}+\Big(((m^{tree}_{\kappa_i^0})^2m_{\kappa_j^0}^{tree}+3m_{\kappa_i^0}^{tree}
(m^{tree}_{\kappa_j^0})^2)
A_+^{({\tiny H_{\beta}^0, \kappa_{\alpha}^0})}
B_-^{({\tiny H_{\beta}^0, \kappa_{\alpha}^0})}
\nonumber \\
&&\hspace{2.cm}
+(3(m^{tree}_{\kappa_i^0})^2m_{\kappa_j^0}^{tree}+m_{\kappa_i^0}^{tree}(m^{tree}_{\kappa_j^0})^2)
A_-^{({\tiny H_{\beta}^0, \kappa_{\alpha}^0})}
B_+^{({\tiny H_{\beta}^0, \kappa_{\alpha}^0})}
\Big)F_{3b}(m_{\kappa_{\alpha}^0},m_{H_{\beta}^0})\Big\}
\label{cpencor}
\end{eqnarray}
\item The internal particles are CP-odd Higgs bosons $A_\beta^0$ ($\beta=1,2,
\cdots,5$) and neutralinos (neutrinos through mixing) $\kappa_\alpha^0$
($\alpha=1,2,\cdots,7$). The coupling constants are
\begin{eqnarray}
&&A_-^{({\tiny A_\beta^0,\kappa_\alpha^0})}=i\frac{e}{2s_wc_w}
C_{onn}^{\alpha j\beta}\nonumber \\
&&A_+^{({\tiny A_\beta^0,\kappa_\alpha^0})}=-i\frac{e}{2s_wc_w}
C_{onn}^{*\alpha j\beta}\nonumber \\
&&B_-^{({\tiny A_\beta^0,\kappa_\alpha^0})}=i\frac{e}{2s_wc_w}
C_{onn}^{\alpha \beta i}\nonumber \\
&&B_+^{({\tiny A_\beta^0,\kappa_\alpha^0})}=-i\frac{e}{2s_wc_w}
C_{onn}^{*\alpha \beta i}\nonumber \\
&&m_{\tiny S}=m_{A_\beta^0},\hspace{2.0cm}m_f=m_{\kappa_\alpha^0}
\label{cponloop}
\end{eqnarray}
with
\begin{eqnarray}
&&C_{onn}^{\alpha\beta\gamma}=\Big(c_wZ_{N}^{\beta 2}-s_wZ_N^{\beta 1}
\Big)\Big(\sum\limits_{\delta=1}^{5}Z_{O}^{\alpha\delta}Z_N^{\gamma 2+\delta}
-2Z_{O}^{\alpha 2}Z_N^{\gamma 3}\Big)
\label{conn}
\end{eqnarray}
and $Z_O$ is the mixing matrix of CP-odd Higgs bosons and neutral particles.
The CP-odd Higgs-neutralino loop corrections to the mass matrix
are
\begin{eqnarray}
&&\delta m_{ij}^{(A^0,\kappa^0)}=-\sum\limits_{\alpha=1}^7
\sum\limits_{\beta=1}^5
\bigg\{\Big[3m_{\kappa_i^0}^{tree}m_{\kappa_j^0}^{tree}
A_-^{({\tiny A_{\beta}^0, \kappa_{\alpha}^0})}
B_-^{({\tiny A_{\beta}^0, \kappa_{\alpha}^0})}
+((m^{tree}_{\kappa_i^0})^2+(m^{tree}_{\kappa_j^0})^2\nonumber \\
&&\hspace{2.cm}+m_{\kappa_i^0}^{tree}m_{\kappa_j^0}^{tree})
A_+^{({\tiny A_{\beta}^0, \kappa_{\alpha}^0})}
B_+^{({\tiny A_{\beta}^0, \kappa_{\alpha}^0})}
\Big]m_{\kappa_\alpha^-}F_{3a}(m_{\kappa_{\alpha}^0},m_{A_{\beta}^0})
\nonumber \\
&&\hspace{2.cm}+\Big[((m^{tree}_{\kappa_i^0})^2m_{\kappa_j^0}^{tree}+3m_{\kappa_i^0}^{tree}
(m^{tree}_{\kappa_j^0})^2)
A_+^{({\tiny A_{\beta}^0, \kappa_{\alpha}^0})}
B_-^{({\tiny A_{\beta}^0, \kappa_{\alpha}^0})}
\nonumber \\
&&\hspace{2.cm}
+(3(m^{tree}_{\kappa_i^0})^2m_{\kappa_j^0}^{tree}+m_{\kappa_i^0}^{tree}(m^{tree}_{\kappa_j^0})^2)
A_-^{({\tiny A_{\beta}^0, \kappa_{\alpha}^0})}
B_+^{({\tiny A_{\beta}^0, \kappa_{\alpha}^0})}
\Big]F_{3b}(m_{\kappa_{\alpha}^0},m_{A_{\beta}^0})\Big\}.
\label{cponcor}
\end{eqnarray}
\item The internal particles are charged Higgs bosons $H_\beta^+$ ($\beta=1,2,
\cdots,8$) and charginos (charged leptons) $\kappa_\alpha^-$
($\alpha=1,2,\cdots,5$). The coupling constants can be written as
\begin{eqnarray}
&&A_-^{({\tiny H_\beta^+, \kappa_\alpha^-})}=-\frac{e}{s_wc_w}C_{Rnk}^{
*\beta\alpha j}\nonumber \\
&&A_+^{({\tiny H_\beta^+, \kappa_\alpha^-})}=\frac{e}{s_wc_w}C_{Lnk}^{
*\beta\alpha j}\nonumber \\
&&B_-^{({\tiny H_\beta^+, \kappa_\alpha^-})}=\frac{e}{s_wc_w}C_{Lnk}^{
\beta\alpha i}\nonumber \\
&&B_+^{({\tiny H_\beta^+, \kappa_\alpha^-})}=-\frac{e}{s_wc_w}C_{Rnk}^{
\beta\alpha i}\nonumber \\
&&m_{\tiny
S}=m_{H_\beta^-},\hspace{2.0cm}m_f=m_{\kappa_\alpha^-}.
\label{chcloop}
\end{eqnarray}
The symbols $C_{Lnk}$, $C_{Rnk}$ are defined as
\begin{eqnarray}
&&C_{Lnk}^{\alpha\beta\gamma}=\bigg\{\sum\limits_{\delta=2}
Z_C^{\alpha \delta}\Big[\frac{1}{\sqrt{2}}
\Big(c_wZ_-^{\beta \delta}Z_N^{\gamma 2}+s_wZ_-^{\beta \delta}
Z_N^{\gamma 1}\Big)-c_wZ_-^{\beta 1}Z_N^{\gamma (2+\delta)}\Big]
\nonumber \\
&&\hspace{2.0cm}+\sum\limits_{IJK}\frac{\lambda_{IJK}}{2es_wc_w}
\Big[Z_C^{\alpha (5+I)}Z_-^{\beta (2+J)}Z_N^{\gamma (4+K)}
-Z_C^{\alpha (5+I)}Z_-^{\beta (2+K)}Z_N^{\gamma (4+J)}\Big]\bigg\}
\nonumber \\
&&C_{Rnk}^{\alpha\beta\gamma}=\bigg\{Z_C^{\alpha 2}
\Big[\frac{1}{\sqrt{2}}\Big(c_wZ_+^{*\beta 2}Z_N^{*\gamma 2}
+s_wZ_+^{*\beta 2}Z_N^{*\gamma 1}\Big)+c_wZ_+^{*\beta 1}Z_N^{\gamma 3}
\Big]+\sqrt{2}s_w\sum\limits_{\delta=3}^{5}Z_C^{\alpha (5+\delta)}
Z_+^{*\beta (2+\delta)}Z_N^{*\gamma 1}\bigg\}
\nonumber \\
&&\hspace{2.0cm}+\sum\limits_{IJK}\frac{\lambda_{IJK}}{2es_wc_w}
\Big[Z_C^{\alpha (2+I)}Z_-^{*\beta (4+J)}Z_N^{*\gamma (2+K)}
-Z_C^{\alpha (2+J)}Z_-^{*\beta (4+I)}Z_N^{*\gamma (2+K)}\Big]
\label{chkn}
\end{eqnarray}
and $Z_C$ is the mixing matrix of charged Higgs bosons and sleptons.
The Charged Higgs-chargino loop corrections to the mass matrix
are
\begin{eqnarray}
&&\delta m_{ij}^{(H^+,\kappa^-)}=-\sum\limits_{\alpha=1}^7
\sum\limits_{\beta=1}^8
\bigg\{\Big[3m_{\kappa_i^0}^{tree}m_{\kappa_j^0}^{tree}
A_-^{({\tiny H_{\beta}^+, \kappa_{\alpha}^-})}
B_-^{({\tiny H_{\beta}^+, \kappa_{\alpha}^-})}
+(m_{\kappa_i^0}^{tree}m_{\kappa_j^0}^{tree}+(m^{tree}_{\kappa_i^0})^2\nonumber \\
&&\hspace{2.0cm}+(m^{tree}_{\kappa_j^0})^2)A_+^{({\tiny H_{\beta}^+, \kappa_{\alpha}^-})}
B_+^{({\tiny H_{\beta}^+, \kappa_{\alpha}^-})}
\Big]m_{\kappa_\alpha^-}F_{3a}(m_{\kappa_{\alpha}^-},m_{H_{\beta}^+})\nonumber \\
&&\hspace{2.cm}+\Big[((m^{tree}_{\kappa_i^0})^2m_{\kappa_j^0}^{tree}
+3m_{\kappa_i^0}^{tree}(m^{tree}_{\kappa_j^0})^2)
A_+^{({\tiny H_{\beta}^+, \kappa_{\alpha}^-})}
B_-^{({\tiny H_{\beta}^+, \kappa_{\alpha}^-})}
+(3(m^{tree}_{\kappa_i^0})^2m_{\kappa_j^0}^{tree}\nonumber \\
&&\hspace{2.0cm}+m_{\kappa_i^0}^{tree}(m^{tree}_{\kappa_j^0})^2)
A_-^{({\tiny H_{\beta}^+, \kappa_{\alpha}^-})}
B_+^{({\tiny H_{\beta}^+, \kappa_{\alpha}^-})}
\Big]F_{3b}(m_{\kappa_{\alpha}^-},m_{H_{\beta}^+})\Big\}.
\label{chccor}
\end{eqnarray}
\item The internal particles are up-type scalar quarks $\tilde{U}^k_\alpha$
($\alpha=1,2$; k=1, 2, 3.)  and quark $u^k$.
The coupling constants can be written as
\begin{eqnarray}
&&A_-^{({\tiny \tilde{U}^k_\alpha, u^k})}=\frac{e}{\sqrt{2}s_wc_w}
Z_{\tilde{U}^k}^{*\alpha 1}\Big(c_wZ_N^{j 2}+\frac{1}{3}s_wZ_N^{j 1}
\Big)-\frac{em_{u^k}}{2\sqrt{2}s_w\sin\beta m_w}Z_{\tilde{U}^k}^{*\alpha
1}Z_N^{j 3}\nonumber \\
&&A_+^{({\tiny \tilde{U}^k_\alpha, u^k})}=\frac{2\sqrt{2}e}{3c_w}
Z_{\tilde{U}^k}^{*\alpha 2}Z_N^{*j 1}-\frac{em_{u^k}}{2\sqrt{2}s_w\sin\beta m_w}Z_{\tilde{U}^k}^{*\alpha 1}Z_N^{*j 3}\nonumber \\
&&B_-^{({\tiny \tilde{U}^k_\alpha, u^k})}=\frac{2\sqrt{2}e}{3c_w}
Z_{\tilde{U}^k}^{\alpha 2}Z_N^{i 1}-\frac{em_{u^k}}{2\sqrt{2}s_w\sin\beta m_w}Z_{\tilde{U}^k}^{\alpha 1}Z_N^{i 3}\nonumber \\
&&B_+^{({\tiny \tilde{U}^k_\alpha, u^k})}=\frac{e}{\sqrt{2}s_wc_w}
Z_{\tilde{U}^k}^{\alpha 1}\Big(c_wZ_N^{*i 2}+\frac{1}{3}s_wZ_N^{*i 1}
\Big)-\frac{em_{u^k}}{2\sqrt{2}s_w\sin\beta m_w}Z_{\tilde{U}^k}^{\alpha
1}Z_N^{*i 3}\nonumber \\
&&m_{\tiny
S}=m_{\tilde{U}^k_\alpha},\hspace{2.0cm}m_f=m_{u^k},
\label{upqloop}
\end{eqnarray}
where $Z_{\tilde{U}^k}$ denotes the mixing matrix of the k-th left-handed
and right-handed up-type
scalar quarks.
The up-type scalar-quark-quark loop corrections to the mass matrix
are
\begin{eqnarray}
&&\delta m_{ij}^{(\tilde{U},u)}=-\sum\limits_{\alpha=1}^2
\sum\limits_{k=1}^3
\bigg\{\Big[3m_{\kappa_i^0}^{tree}m_{\kappa_j^0}^{tree}
A_-^{({\tiny \tilde{U}^k_\alpha, u^k})}
B_-^{({\tiny \tilde{U}^k_\alpha, u^k})}
+((m^{tree}_{\kappa_i^0})^2+(m^{tree}_{\kappa_j^0})^2\nonumber \\
&&\hspace{2.cm}+m_{\kappa_i^0}^{tree}m_{\kappa_j^0}^{tree})
A_+^{({\tiny \tilde{U}^k_\alpha, u^k})}
B_+^{({\tiny \tilde{U}^k_\alpha, u^k})}
\Big]m_{u^k}F_{3a}(m_{u^k},m_{\tilde{U}^k_\alpha})\nonumber \\
&&\hspace{2.cm}+\Big[((m^{tree}_{\kappa_i^0})^2m_{\kappa_j^0}^{tree}
+3m_{\kappa_i^0}^{tree}(m^{tree}_{\kappa_j^0})^2)
A_+^{({\tiny \tilde{U}^k_\alpha, u^k})}
B_-^{({\tiny \tilde{U}^k_\alpha, u^k})}
+(m_{\kappa_i^0}^{tree}(m^{tree}_{\kappa_j^0})^2\nonumber \\
&&\hspace{2.cm}+3(m^{tree}_{\kappa_i^0})^2m_{\kappa_j^0}^{tree})
A_-^{({\tiny \tilde{U}^k_\alpha, u^k})}
B_+^{({\tiny \tilde{U}^k_\alpha, u^k})}
\Big]F_{3b}(m_{u^k},m_{\tilde{U}^k_\alpha})\Big\}
\label{upqcor}
\end{eqnarray}
\item The internal particles are down-type scalar quarks $\tilde{D}^k_\alpha$
($\alpha=1,2$; k=1, 2, 3.)  and quark $d^k$.
The coupling constants are written as
\begin{eqnarray}
&&A_-^{({\tiny \tilde{D}^k_\alpha, d^k})}=\frac{e}{\sqrt{2}s_wc_w}
Z_{\tilde{D}^k}^{*\alpha 1}\Big(-c_wZ_N^{j 2}+\frac{1}{3}s_wZ_N^{j 1}
\Big)+\frac{1}{2}\sum\limits_{K=0}^3\lambda_{d}^{Kkk}
Z_{\tilde{D}^k}^{*\alpha 2}Z_N^{j (4+K)},\nonumber \\
&&A_+^{({\tiny \tilde{D}^k_\alpha, d^k})}=-\frac{\sqrt{2}e}{3c_w}
Z_{\tilde{D}^k}^{*\alpha 2}Z_N^{*j 1}
+\frac{1}{2}\sum\limits_{K=0}^3\lambda_{d}^{Kkk}
Z_{\tilde{D}^k}^{*\alpha 1}Z_N^{*j (4+K)},\nonumber \\
&&B_+^{({\tiny \tilde{D}^k_\alpha, d^k})}=-\frac{\sqrt{2}e}{3c_w}
Z_{\tilde{D}^k}^{\alpha 2}Z_N^{i 1}
+\frac{1}{2}\sum\limits_{K=0}^3\lambda_{d}^{Kkk}
Z_{\tilde{D}^k}^{\alpha 1}Z_N^{i (4+K)},\nonumber \\
&&B_-^{({\tiny \tilde{D}^k_\alpha, d^k})}=\frac{e}{\sqrt{2}s_wc_w}
Z_{\tilde{D}^k}^{\alpha 1}\Big(-c_wZ_N^{*i 2}+\frac{1}{3}s_wZ_N^{*i 1}
\Big)+\frac{1}{2}\sum\limits_{K=0}^3\lambda_{d}^{Kkk}
Z_{\tilde{D}^k}^{\alpha 2}Z_N^{i (4+K)}\nonumber \\
&&m_{\tiny S}=m_{\tilde{D}^k_\alpha},\hspace{2.0cm}m_f=m_{d^k},
\label{dwqloop}
\end{eqnarray}
where $Z_{\tilde{D}^k}$ denotes the mixing matrix of k-th left-handed
and right-handed down-type scalar quarks.
The up type scalar quark-quark loop corrections to the mass matrix
are
\begin{eqnarray}
&&\delta m_{ij}^{(\tilde{D},d)}=-\sum\limits_{\alpha=1}^2
\sum\limits_{k=1}^3
\bigg\{\Big[3m_{\kappa_i^0}^{tree}m_{\kappa_j^0}^{tree}
A_-^{({\tiny \tilde{D}^k_\alpha, d^k})}
B_-^{({\tiny \tilde{D}^k_\alpha, d^k})}
+((m^{tree}_{\kappa_i^0})^2+(m^{tree}_{\kappa_j^0})^2\nonumber \\
&&\hspace{2.cm}+m_{\kappa_i^0}^{tree}m_{\kappa_j^0}^{tree})
A_+^{({\tiny \tilde{D}^k_\alpha, d^k})}
B_+^{({\tiny \tilde{D}^k_\alpha, d^k})}
\Big]m_{d^k}F_{3a}(m_{d^k},m_{\tilde{D}^k_\alpha})\nonumber \\
&&\hspace{2.cm}+\Big[(3(m^{tree}_{\kappa_i^0})^2m_{\kappa_j^0}^{tree}
+m_{\kappa_i^0}^{tree}(m^{tree}_{\kappa_j^0})^2)
A_+^{({\tiny \tilde{D}^k_\alpha, d^k})}
B_-^{({\tiny \tilde{D}^k_\alpha, d^k})}
+((m^{tree}_{\kappa_i^0})^2m_{\kappa_j^0}^{tree}\nonumber \\
&&\hspace{2.cm}+3m_{\kappa_i^0}^{tree}(m^{tree}_{\kappa_j^0})^2)
A_-^{({\tiny \tilde{D}^k_\alpha, d^k})}
B_+^{({\tiny \tilde{D}^k_\alpha, d^k})}
\Big]F_{3b}(m_{d^k},m_{\tilde{D}^k_\alpha})\Big\}.
\label{dwqcor}
\end{eqnarray}
\end{itemize}

\begin{figure}
\setlength{\unitlength}{1mm}
\begin{picture}(230,205)(55,90)
\put(30,0){\includegraphics{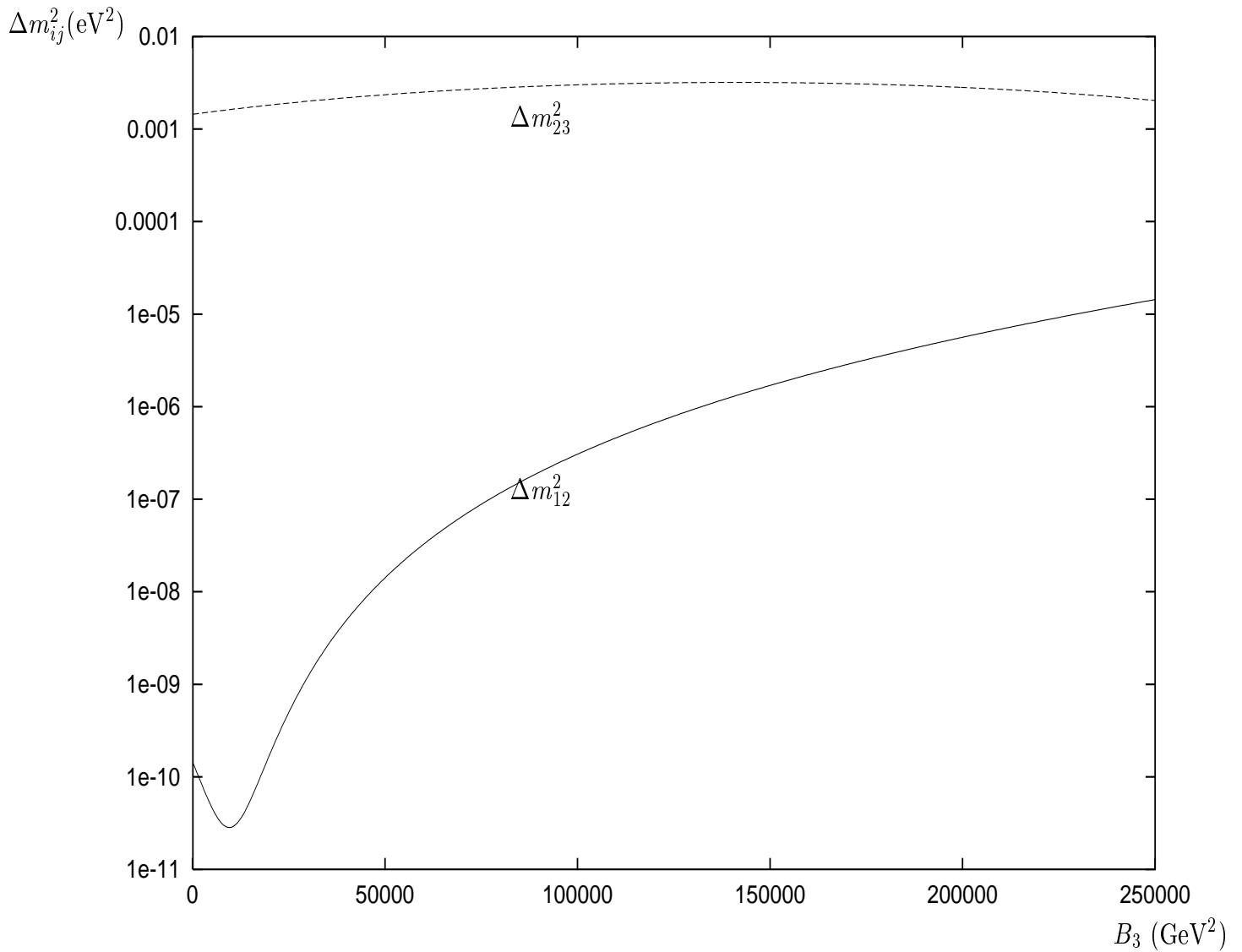}}
\end{picture}
\caption[]{Dependence of $\Delta m^2_{ij}$ on the soft R-parity violation
parameter $B^3$ with (a)solid-line: $\Delta m^2_{12}$
and (b)dash-line: $\Delta m^2_{23}$. The X-axis corresponds to
the $B^3$ (in GeV$^2$) and the Y-axis corresponds to the $\Delta m^2_{ij}$
(in eV$^2$). The other parameters are taken as in the text.}
\label{fig1}
\end{figure}
\end{document}